\documentclass[11pt,dvips]{article}
\usepackage{graphicx}
\usepackage{amsmath}
\usepackage{amssymb}
\usepackage{latexsym}
\usepackage{color}
\begin{document}
\thispagestyle{empty}
\begin{center}
\vspace{1.8cm}
{\Large{\bf Unified scheme for correlations using linear relative entropy }}\\

\vspace{1.5cm} {\bf M. Daoud}$^{a,b,c}${\footnote { email: {\sf
m$_{-}$daoud@hotmail.com}}}, {\bf R. Ahl Laamara}$^{d,e}$ {\footnote
{ email: {\sf ahllaamara@gmail.com}}}  and {\bf W. Kaydi}$^{d}$
{\footnote { email: {\sf kaydi.smp@gmail.com}}} \\
\vspace{0.5cm}
$^{a}${\it Max Planck Institute for the Physics of Complex Systems, Dresden, Germany}\\[0.5em]
$^{b}${\it Abdus Salam  International Centre for Theoretical Physics, Trieste, Italy}\\[0.5em]
$^{c}${\it Department of Physics, Faculty of Sciences,  University Ibnou Zohr, Agadir , Morocco}\\[0.5em]
$^{d}${\it LPHE-Modeling and Simulation, Faculty  of Sciences, Rabat, Morocco}\\[0.5em]
$^{e}${\it Centre of Physics and Mathematics, Rabat, Morocco}\\[0.5em]
\vspace{3cm} {\bf Abstract}
\end{center}
\baselineskip=18pt
\medskip

A linearized variant of relative entropy is used to quantify in a
unified scheme the different kinds of correlations in a bipartite
quantum system. As illustration, we consider a two-qubit state with
parity and exchange symmetries  for which we determine the total,
classical and quantum correlations. We also give the explicit expressions
of its closest product state,  closest classical state and the
corresponding  closest product state.  A closed additive
relation, involving the various correlations quantified  by linear
relative entropy, is derived.

\newpage
\section{Introduction}

Quantum entanglement in quantum systems, comprising two or more
parts, constitutes a key concept to distinguish between quantum and
classical correlations  and subsequently to understand
quantum-classical boundary. Besides its fundamental
aspects, entanglement is commonly accepted of paramount importance
 in the development of quantum information science
\cite{NC-QIQC-2000,Vedral-RMP-2002,
Horodecki-RMP-2009,Guhne,Amico,Modi1}. In fact, entangled states have found
various applications in quantum information processing protocols as for
instance quantum cryptography \cite{Eckert}, quantum teleportation
\cite{Ben1}, quantum dense coding \cite{Ben2}. Nowadays,
entanglement is recognized as a valuable resource in several communication and computational tasks
\cite{Fuchs,Rausschendorf,Gottesman}. In view of these remarkable
realizations and implementations, the concept of entanglement is
expected to have many other implications and applications in others
areas of research, especially condensed matter physics.

Therefore, the quantification and the characterization of quantum
correlations between the components of a composite quantum system
have attracted a special attention during the last two decades. The
experimental and theoretical efforts, deployed in this context, are
essential to develop the appropriate strategies to prevent against
the decoherence effects induced by the system-environment coupling
(see for instance the recent works \cite{Xu,LoFranco,Silva} and
references therein). Different measures were introduced from
different perspectives and for various purposes
\cite{Rungta,Ben3,Wootters,Coffman,Aaronson0,PalulaFM,Bromley}.
Probably the most familiar among them is the quantum discord
\cite{Vedral-et-al,Ollivier-PRL88-2001} which goes beyond the
entanglement of formation  \cite{Wootters98,Hil97}. It is given by
the difference of total and classical correlations existing in a
bipartite system. Now, it is well understood that almost all quantum
states, including unentangled (separable) ones, possess quantum
correlations. However, the analytical evaluation of quantum discord
requires extremization procedures that can be tedious to achieve
\cite{Luo,Ali,Shi1,Shi2,Girolami,Fanchini,Giorda,Adesso1}. To
overcome this difficulty, a geometrical approach was proposed in
\cite{Dakic2010}. It is based on the Hilbert-Schmidt norm in the
space of density matrices. This measure   provides explicit
analytical expressions for pairwise quantum correlations. Clearly,
Hilbert-Schmidt norm is not the unique distance which can be defined
in the space of quantum states. Several distances are possible
(trace distance,  Bures distance, ...) with their own advantages and
drawbacks and each one might be useful for some appropriate purpose
\cite{Dajka,Paula1,Aaronson,Paula2}.

The states of any multipartite quantum system can be classified as
being classical, quantum-classical and quantum states. Subsequently,
the correlations can also be categorized in total, quantum,
semi-classical (related to quantum-classical states) and classical
correlations. This classification requires a specific measure
(entropic or geometric distance) to decide about the dissimilarity
between a given quantum state and its closest one without the
desired property and to provide a consistent scheme to treat equally
the different correlations. In this sense, using the relative
entropy, an approach unifying the correlations in multipartite
systems was recently developed in \cite{Modi}. In particular,   a
very significant and interesting  additivity relation was  reported
($ D+C = T+L $). It states that the sum of quantum  $D$ and
classical $C$ correlations is equal to the sum of  total mutual
correlations $T$ and another quantity $L$ that is exactly the
difference between $D$ and the quantum discord as originally
introduced in \cite{Vedral-et-al,Ollivier-PRL88-2001}.

However, it must be noticed that, despite its theoretical
information meaning,  the relative entropy  is not symmetric in its
arguments and therefore can not be considered as a true metric
distance. In the other hand, from an analytical point of view, the
derivation of closed expressions of relative entropy based measures
involves optimization procedures that are in general very
complicated to perform. In this respect, a purely geometrical
unified framework to classify the correlations in a given quantum
state was discussed in \cite{Bellomo1,Bellomo2}. Using the
Hilbert-Schmidt norm and paralleling the definition of the geometric
discord, the geometric measures of total and classical correlations
in a two qubit system were derived in \cite{Bellomo1,Bellomo2}. In
contrast with the relative entropy, the additivity relation of the
type ($ D+C = T+L $) is not, in general, satisfied.

In this paper, we introduce a linearized variant of relative
entropy. We obtain the explicit analytical expressions of quantum
and classical correlations in a two qubit system. The relation with
the geometric measure based on Hilbert-Schmidt norm is established.
We show that the linear relative entropy provides us with a simple
approach to treat the different kinds of bipartite correlations in a
common framework. This approach can be seen, in some sense,
interpolating between the relative entropy-based
\cite{Vedral-et-al,Ollivier-PRL88-2001}
 and Hilbert-Schmidt-based \cite{Bellomo1,Bellomo2}  classification schemes. More specifically,
it provides us with a very simple way to perform the optimizations
required in deriving closest product, classical and classical
product states. We also show that the correlations satisfy a closed
additivity.

 This paper is organized as follows. In section 2,
we decompose the linear entropy in symmetric and anti-symmetric parts.  We show that the
 antisymmetric part is related to quantum Jensen-Shannon divergence and
the symmetric part is exactly the Hilbert-Schmidt
 distance {\cite{Bellomo1}. Using the linear
 relative entropy,
 we obtain a closed additivity relation of
 the various bipartite correlations existing in a two qubit system. A comparison with Hilbert-Schmidt
 based approach is also investigated. As illustration, we consider, in section 3, a bipartite system
possessing the parity
 symmetry and invariant under qubits permutation. In this situation, the explicit derivations of the suitable closest
product and classical states is achieved.
  The analytical expressions of total, quantum and classical
 correlations are obtained and the additivity relation is
 discussed. Concluding remarks close this paper.

\section{ Correlation quantifiers based on symmetrized linear relative entropy}

\subsection{Correlation quantifiers based on relative entropy}

The relative entropy offers the appropriate scheme to unify
the different kinds of correlations existing
in multipartite systems
\cite{Modi}. It is the quantum analogue of the Kullback-Leibler
divergence between two classical probability distributions and
characterizes  the dissimilarity  between two quantum states.
The relative entropy  defined by
\begin{equation}\label{RE}
S(\rho\|\sigma)=-\mathrm{Tr}(\rho\log\sigma)-S(\rho),
\end{equation}
constitutes  a quantitative tool to distinguish between the states
of a given degree of quantumness and gives the distance between them
according to the nature of their properties
($S(\rho)=-\mathrm{Tr}(\rho\log\rho)$ is the von Neumann entropy).
For a bipartite system, the total correlation
$T=S(\rho\|\pi_{\rho})$ is quantified by the relative entropy
between a state
 $\rho$ and its  closest product
state $\pi_{\rho}=\rho_A\otimes\rho_B$, where $\rho_A$ and $\rho_B$
denote the reduced density matrices of the subsystems. It writes as
the difference of the von Neumann entropies \cite{Modi}
\begin{equation}\label{T}
T=S(\rho\|\pi_{\rho}) = S(\pi_{\rho}) - S(\rho).
\end{equation}
 Similarly, the quantum discord, which encompasses  quantum correlations, is measured as  the minimal distance between the state
$\rho$ and its closest classical state
\begin{equation}\label{chirho}
\chi_\rho=\sum_{i,j}p_{i,j}\vert{i}\rangle\langle{i}\vert\otimes\vert{j}\rangle
\langle{j}\vert,
\end{equation}
where  $p_{i,j}$ are the probabilities and
$\{\vert{i}\rangle,\vert{j}\rangle\}$ local basis. It writes also as
the difference between the von Neumann entropies of the states
$\rho$ and $\chi_{\rho}$ \cite{Modi}
\begin{equation}\label{D}
D=S(\rho\|\chi_{\rho}) =  S(\chi_{\rho}) - S(\rho).
\end{equation}
The classical correlation  gives the distance between the closest
 classical state $\chi_{\rho}$ and its
closest classical product state $\pi_{\chi_{\rho}}$.  It coincides with the difference of von Neumann entropies of the relevant states
\begin{equation}\label{C}
 C=S(\chi_{\rho}\|\pi_{\chi_{\rho}}) = S(\pi_{\chi_{\rho}}) - S(\chi_{\rho}).
\end{equation}
In this approach the relative entropy-based quantum correlations or
quantum discord $D$ (\ref{D}) does not coincide with the original
definition of discord introduced in
\cite{Vedral-et-al,Ollivier-PRL88-2001}. The difference is given by
\cite{Modi}
\begin{equation}\label{L}
 L=S(\pi_{\rho}\|\pi_{\chi_{\rho}}) = S(\pi_{\chi_{\rho}}) - S(\pi_{\rho}).
\end{equation}
The  entropy-based correlations $T$, $D$, $C$ and $L$ are expressed as  differences of  von Neumann entropies
(Eqs. (\ref{T}), (\ref{D}), (\ref{C}) and (\ref{L})) and they satisfy   the
following remarkable additivity relation \cite{Modi}
\begin{equation}\label{additive relation RE}
 T- D- C + L = 0.
\end{equation}
It must be emphasized that the relative entropy  (\ref{RE}) is not
symmetric under the exchange $\rho \leftrightarrow \sigma$. In this
respect, it cannot define a distance from a purely mathematical
point of view. Moreover, as mentioned above, the analytical evaluation of relative entropy-based correlations  requires intractable
minimization procedures. To avoid this problem,  the linear relative entropy
offers an alternative way to get computable expressions of
correlations existing in multipartite systems \cite{Bellomo1}.

\subsection{ Linear relative entropy}

The linear entropy
$$ S_2(\rho) \, \dot= \, 1 - \mbox{Tr}(\rho^2)
\, $$ is related to the degree of purity, $P = \mbox{Tr}(\rho^2)$,
and therefore reflects the mixedness in the state $\rho$. It is
defined as a linearized variant of von Neumann entropy by
approximating $\log \rho$ by  $  \rho - \mathbb{I}$ where
$\mathbb{I}$ stands for the identity matrix. Accordingly, the
relative entropy (\ref{RE}) can be linearized  as follows
{\cite{Bellomo1}
\begin{equation}\label{lre}
S_l(\rho_1 \|\rho_2) = {\rm Tr} \rho_1(\rho_1 - \rho_2).
\end{equation}
It is not symmetric under the interchange of the states $\rho_1$ and
$\rho_2$. To define a symmetrized linear relative entropy,
$S_l(\rho_1 \|\rho_2)$ is decomposed as the sum of two terms:
symmetric and antisymmetric. The symmetric part is defined by
\begin{equation}\label{s+}
S_+(\rho_1 \|\rho_2)  =   S_l(\rho_1 \|\rho_2) +  S_l(\rho_2
\|\rho_1).
\end{equation}
The antisymmetric term is given by
\begin{equation}\label{s-}
S_-(\rho_1 \|\rho_2)  =   S_l(\rho_1 \|\rho_2) -  S_l(\rho_2
\|\rho_1)
\end{equation}
and rewrites as the differences between the linear entropies of the states $\rho_1$ and $\rho_2$
\begin{equation}\label{s-2}
S_-(\rho_1 \|\rho_2)  =   S_2(\rho_2) -  S_2(\rho_1).
\end{equation}
We emphasize that the antisymmetric linear relative entropy
(\ref{s-}) is related  to the quantum Jensen-Shannon entropy of
order 2 defined by
\begin{equation}\label{JS}
D_2(\rho_1 \|\rho_2) :=  S_2 \bigg(\frac{\rho_1 + \rho_2}{2}\bigg) -
\frac{1}{2} S_2(\rho_1)- \frac{1}{2} S_2(\rho_2).
\end{equation}
 Indeed, it can be
expressed as
\begin{equation}
S_-(\rho_1 \|\rho_2) =   D_2(\rho_1 + \rho_2  \|\rho_2 - \rho_1 ) - D_2(\rho_1 + \rho_2  \|\rho_1- \rho_2 ).
\end{equation}
The quantum Jensen-Shannon entropy of order 2 (\ref{JS}) is a
symmetrized form of the linear relative entropy. It was recently
used to investigate the distance between between any two density
operators
 (see  for instance \cite{Briet,Lamberti} and references quoted therein) and
subsequently constitutes an adequate geometric tool to classify
quantum states according to their properties. The square root of
quantum Jensen-Shannon divergence is a metric and can be
isometrically embedded in a real Hilbert space equipped with a
Hilbert-Schmidt norm \cite{Briet}. This result is useful for our
purpose. In fact, using (\ref{lre}) and (\ref{s+}), it is simple to
check that the symmetric part of linear relative entropy is exactly
the Hilbert-Schmidt distance  \cite{Bellomo1}
 \begin{equation}\label{s+2}
S_+(\rho_1 \|\rho_2)  =  \|\rho_1 - \rho_2\|^2.
\end{equation}
The symmetric (\ref{s+})  and antisymmetric (\ref{s-2}) linear
entropy are the main ingredients in this work. The symmetric linear
relative entropy measures
 the distance between the states of a given
quantum system and the antisymmetric linear relative entropy
quantifies the amount of correlations existing between two distinct
states.  With the linear relative entropy, considerable
simplifications are possible in deriving the   explicit expressions
of total, quantum and classical pairwise correlations.
 Furthermore,  interesting relations between the correlations (as
measured by linear relative entropy) and their geometric
counterparts (as defined in \cite{Bellomo1,Bellomo2} using the Hilbert-Schmidt distance) can be
derived. This issue is discussed, in the next subsection, for an arbitrary two qubit system.

\subsection{ Additivity relation of geometric and entropic correlations}

The Fano-Bloch representation of an arbitrary two-qubit state $\rho$ is
\begin{eqnarray}\label{block-rep}
  \rho & = & \frac{1}{4} \sum_{\alpha,\beta} ~ R_{\alpha \beta} ~\sigma_{\alpha}\otimes\sigma_{\beta}
\end{eqnarray}
where $\alpha, \beta = 0, 1, 2, 3$, $R_{i0} = {\rm
Tr}\rho(\sigma_{i}\otimes \sigma_{0}),~  R_{0i} = {\rm
Tr}\rho(\sigma_{0}\otimes\sigma_{i})$ are components of local Bloch
vectors and $R_{ij} = {\rm Tr}\rho(\sigma_{i}\otimes\sigma_{j})$ are
components of the correlation tensor.  The operators $\sigma_i$ $( i
= 1, 2, 3)$ stand for the three Pauli matrices and $\sigma_0$ is the
identity matrix. The distance (\ref{s+2}), between two distinct
density matrices $\rho$ and $ \rho'$, writes as
\begin{eqnarray}\label{distance-def}
 S_+(\rho \|\rho') \equiv  d(\rho, \rho') & = &  \frac{1}{4} \sum_{\alpha, \beta} ~ (R_{\alpha \beta} -
 R'_{\alpha \beta})^2,
\end{eqnarray}
 in terms of the Fano-Bloch coefficients $R_{\alpha \beta}$ (resp. $R'_{\alpha \beta}$)
 corresponding to $\rho$ (resp. $\rho'$).
The linear analogues  of total correlation $T$ (\ref{T}), quantum correlation
$D$ (\ref{D}), classical correlation $C$ (\ref{C}) and the quantity
$L$ (\ref{L}) are respectively given by
\begin{eqnarray}\label{T2}
T_2 =  S_-(\rho \| \pi_\rho)\qquad D_2 =  S_-(\rho \| \chi_\rho) \qquad C_2 = S_-(\chi_\rho\| \pi_{\chi_\rho}) \qquad L_2 = S_-(\pi_\rho\| \pi_{\chi_\rho}).
\end{eqnarray}
Using the formula (\ref{s-2}), it is easy to see that the
correlations $T_2$, $D_2$, $D_2$ and $L_2$ can be written as
differences of linear entropies. This implies the remarkable
additivity relation
\begin{eqnarray}\label{additivity 2}
T_2 - D_2 -C_2 + L_2 = 0.
\end{eqnarray}
Inspired by the definition of the geometric discord based on
 Hilbert-Schmidt distance \cite{Dakic2010},  Bellomo et al investigated
a geometrical  unified scheme in which
the total correlation $T$ (\ref{T}), the quantum correlation
$D$ (\ref{D}), the classical correlation $C$ (\ref{C}) and the quantity
$L$ (\ref{L})  are redefined
as \cite{Bellomo2}
\begin{equation}\label{geometric total}
    T_\mathrm{g} \equiv\|\rho-\pi_\rho\|^2, \qquad
C_\mathrm{g} \equiv\|\chi_\rho-\pi_{\chi_\rho}\|^2, \qquad
D_\mathrm{g}(\rho)=\|\rho-\chi_\rho\|^2, \qquad L_\mathrm{g}
\equiv\|\pi_\rho-\pi_{\chi_\rho}\|^2.
\end{equation}
It is worthwhile to mention that  in view of the relation between
the distance (\ref{s+}) and the Hilbert-Schmidt norm given by
(\ref{s+2}), the linear relative entropy based  correlations can be
expressed in terms of their geometric counterparts. Indeed, using
(\ref{T2}), one shows
\begin{equation}\label{geometric total-1}
  T_2 =  T_\mathrm{g} - 2  S_2(\pi_\rho \| \rho), \quad   D_2 = D_\mathrm{g}- 2 S_2(\chi_\rho \| \rho) , \quad
C_2 = C_\mathrm{g}  -  2 S_2(\pi_{\chi_\rho} \| \chi_{\rho}), \quad L_2 = L_\mathrm{g} -  2 S_2(\pi_{\chi_\rho} \| \pi_{\rho})
\end{equation}
or alternatively
\begin{equation}\label{geometric total-2}
  T_2 =  T_\mathrm{g} + 2  {\rm Tr} \big(\pi_\rho (\rho - \pi_\rho) \big), ~   D_2 = D_\mathrm{g}+2 {\rm Tr} \big(\chi_\rho (\rho - \chi_\rho )\big) , ~
C_2 = C_\mathrm{g}  + 2 {\rm Tr} \big(\pi_{\chi_\rho}  (\chi_{\rho} - \pi_{\chi_\rho})\big), ~ L_2 = L_\mathrm{g} - 2 {\rm Tr} \big(\pi_{\chi_\rho} (\pi_{\rho} - \pi_{\chi_\rho})\big)
\end{equation}
Reporting the equations (\ref{geometric total-1}), or equivalently (\ref{geometric total-2}), in
the additivity relation (\ref{additivity 2}), one verifies that the geometric correlations (\ref{geometric total}) satisfy the relation
\begin{equation}\label{geometric additivity}
T_\mathrm{g} -  D_\mathrm{g}- C_\mathrm{g} + L_\mathrm{g} = \Delta_\mathrm{g}
\end{equation}
where
\begin{equation}\label{deltag}
 \Delta_\mathrm{g} = 2 \bigg[ {\rm Tr}\big( \pi_{\rho} (\pi{_\rho} - \rho) \big) + {\rm Tr}\big( \pi_{\chi_{\rho}} ( \chi_{\rho} - \pi_{\rho}) \big) \bigg].
\end{equation}
We notice that the classical states $\chi_\rho$ (\ref{chirho}) satisfy the relation
${\rm Tr} \rho \chi_\rho = {\rm Tr} \chi^2_\rho $ \cite{Dakic2010}. Hence,  the geometric discord $D_\mathrm{g}$ coincides with
the linear quantum correlation $D_2$ ($D_2 = D_\mathrm{g} $). The geometric correlations satisfy
 a closed additivity relation similar to (\ref{additivity 2}) when the quantity $ \Delta_\mathrm{g} $  (\ref{deltag}) is zero.
A detailed  analysis of  the relationship among  the geometric correlations
$T_\mathrm{g}$,  $D_\mathrm{g}$, $C_\mathrm{g}$ and $L_\mathrm{g}$
for two qubit $X$ states was recently investigated  in \cite{Bellomo2}. In particular, it has been shown that the quantity  $ \Delta_\mathrm{g} $
vanishes only in some special cases  like for instance
Bell states \cite{Bellomo1}. In this respect, to understand the differences
between linear relative entropy  and Hilbert-Schmidt based quantifiers, we
shall consider a specific class of two qubit states for which one can explicitly
evaluated total, quantum and classical pairwise correlations.  This constitutes
the main objective of the next section.

\section{Analytical expressions of correlations }
To illustrate the general results discussed in the previous section, we shall consider a family of two qubit
density matrices whose entries are specified in terms of two real
parameters. They are defined as
\begin{equation}\label{rho}
  \rho= \left(%
\begin{array}{cccc}
  c_1& 0 & 0 &  \sqrt{c_1c_2}  \\
  0 &  \frac{1}{2}(1-c_1-c_2) & \frac{1}{2}(1-c_1-c_2) & 0 \\
  0 &  \frac{1}{2}(1-c_1-c_2) &  \frac{1}{2}(1-c_1-c_2) & 0 \\
   \sqrt{c_1c_2}  & 0 & 0 &  c_2 \\
\end{array}%
\right)
\end{equation}
in the computational  basis $\mathcal{B}=\{\vert{00}\rangle ,
\vert{01}\rangle, \vert{10}\rangle, \vert{11}\rangle\}.$ The
parameters $c_1$ and $c_2$ satisfy the conditions $0 \leq c_1 \leq
1$ , $0 \leq c_2 \leq 1$ and  $ 0 \leq c_1 + c_2 \leq 1$. We have
taken all entries positives. In fact, the local unitary
transformation
$$ \vert 0 \rangle_k \rightarrow \exp \bigg(\frac{i}{2} (\theta_{1} + (-)^k\theta_{2})\bigg) \vert 0 \rangle_k $$
eliminates the phase factors of the off diagonal elements and  the rank of the density matrix $\rho$ remains unchanged. In
the Fano-Bloch representation, the density $\rho$  takes the form (\ref{block-rep}) and the corresponding (non-vanishing)
 correlation  matrix elements are
\begin{eqnarray}\label{R3}
R_{30} =  R_{03} =  c_1 - c_2 \qquad R_{33} = 2(c_1 + c_2)-1,
\end{eqnarray}
\begin{eqnarray}\label{R12}
R_{11} =  1 -  (\sqrt{c_1}- \sqrt{c_2})^2 \qquad R_{22} =  1 -
(\sqrt{c_1}+ \sqrt{c_2})^2.
\end{eqnarray}
The density matrix (\ref{rho}) is invariant under parity symmetry and exchange
transformation ($\rho$ commutes with $\sigma_3 \otimes \sigma_3$ and
the permutation operator which exchanges the qubit state $\vert i ,
j \rangle$ to  $\vert j , i \rangle$ leaves $\rho$ unchanged). These symmetries  simplify considerably the complexity of the
analytical evaluations of bipartite correlations.


\subsection{Total correlation and closest product state}\label{subsect-prho}
\subsection{Closest product state}
Let us derive the explicit expression of the total
of total correlation $T_2$ defined by (\ref{T2}) in the bipartite state $\rho$
(\ref{rho}). For this end, we first
determine the closest product state to the density matrix $\rho$
(\ref{rho}).  An arbitrary product state  writes
\begin{eqnarray}\label{productrho}
  \pi_{\rho}  =  \rho_1 \otimes  \rho_2  = \frac{1}{4}\bigg[ \sigma_{0}\otimes \sigma_{0} +\sum_{i=1}^{3}(a_{i}\sigma_{i}\otimes \sigma_{0}
   +b_{i} \sigma_{0}\otimes\sigma_{i})+\sum_{i,j=1}^{3} a_{i}b_{j}  \sigma_{i}\otimes\sigma_{j}\bigg]
\end{eqnarray}
where $\vec{a} = (a_1,a_2,a_3)$ and $\vec{b} = (b_1,b_2,b_3)$ denote the unit Bloch vectors of the states $\rho_1$ and $\rho_2$
\begin{eqnarray}\label{rho1rho2}
 \rho_1 & = & \frac{1}{2} \bigg[ \sigma_{0} + \sum_{i=1}^{3} a_{i}\sigma_{i} \bigg], \qquad  \rho_2 ~ = ~ \frac{1}{2}
\bigg[ \sigma_{0}  +\sum_{i=1}^{3}  b_{i}\sigma_{i} \bigg].
\end{eqnarray}
Due to the symmetry of the state $\rho$  under exchange of the qubits 1 and 2, the product
state $\pi_{\rho}$ is also symmetric.  This implies
$$ a_i = b_i \qquad i = 1,2,3.$$
Furthermore, the parity symmetry of the density matrix $\rho$ $(~[
\rho , \sigma_3 \otimes \sigma_3] = 0)$ implies the
parity invariance of the state $\pi_{\rho}$ (\ref{productrho}). This  imposes
$$ a_i = b_i = 0 \qquad i = 1,2.$$
Using the definition (\ref{distance-def}), the distance between the
state $\rho$ and $\pi_{\rho}$ takes the simple form
\begin{eqnarray}\label{d}
  d(\rho, \pi_{\rho}) & = & \frac{1}{4} \big[2(R_{30} - a_3)^2  + R_{11}^2 + R_{22}^2 + (R_{33} - a_3^2)^2\big]
\end{eqnarray}
to be optimized with respect one variable only, i.e. $a_3$. It is clear that the
parity and exchange symmetries simplify the minimization process in
determining  the closest product state. A
minimal  distance (\ref{d}) is obtained   for  $a_3$ satisfying the  following equation
\begin{eqnarray}\label{cubic-equation}
a_3^3  + a_3 (1 - R_{33}) - R_{30}  = 0.
\end{eqnarray}
\noindent  This cubic equation can be solved using Cardano's
formula. Here, it is necessary to stress
 that for $X$ states without exchange symmetry, the explicit
determination of the Bloch coefficients $a_i$ and $b_i$  in
(\ref{rho1rho2}) is complicated (see \cite{Bellomo2}). This explains
why we deliberately chose to consider only  two qubit states of type
(\ref{rho}) which have  both permutation and parity symmetries.

\noindent Being constrained to real solutions, the only real solution of (\ref{cubic-equation}) is
\begin{eqnarray}\label{solutiona3}
a_3 = \sqrt[3]{\frac{\sqrt{\Delta} + R_{30}}{2}}
 - \sqrt[3]{\frac{ \sqrt{\Delta} - R_{30}}{2}}
\end{eqnarray}
where the discriminant $\Delta$ given by
$$\Delta = R_{30}^2 + \frac{4}{27}(1 - R_{33})^3$$
is always positive ( $R_{33} \leq 1$). It follows that the closest product state to $\rho$ (\ref{productrho}) takes the form
\begin{eqnarray}\label{pirho}
  \pi_{\rho} & = & \frac{1}{4}\bigg[ \sigma_{0}\otimes \sigma_{0} + a_{3}~\sigma_{3}\otimes \sigma_{0}
   +a_{3}~ \sigma_{0}\otimes\sigma_{3} +  a^2_{3} ~  \sigma_{3}\otimes\sigma_{3}\bigg].
\end{eqnarray}
It is interesting to note that for $c_1=c_2$, the density matrix
$\rho$ (\ref{rho}) becomes a Bell-diagonal state.  In this special
case, the matrix elements $R_{30}$ and $R_{03}$, given by
(\ref{R3}), vanish and from (\ref{solutiona3}) we have $a_3 = 0$.
This implies that the closest product state (32) is just the product
of the marginals $\rho_1 = \frac{1}{2}\sigma_0$ and $\rho_2 =
\frac{1}{2}\sigma_0$ of $\rho$.

\subsubsection{Total correlation}
Using (\ref{T2}) together with (\ref{s-}), the total correlation writes
\begin{equation}\label{T2c1c2}
T_2 =\frac{1}{4}\big[ 2 ( R^2_{03} - a^2_3 ) + R^2_{11} + R^2_{22} + (R^2_{33}-a^4_3)].
\end{equation}
The behavior of total correlation $T_2$  versus $c_1$ is given in
the figure 1 for different values of $\alpha = c_1+c_2$ $(\alpha =
0.1, 0.2 \cdots, 0.9)$. In this figure, as well as in the other figures
presented in this paper, the parameter $c_1$ vary  from 0 to
$\alpha$.
The minimal value of the total
correlation is obtained for $(c_1 = 0, c_2 = \alpha)$ and $(c_1 =
\alpha,c_2=0)$. These two situations correspond respectively to
states of the form
\begin{equation}\label{c1=0,c2=alpha}
\rho(c_1 = 0, c_2 = \alpha) = \alpha \vert 11 \rangle \langle 11
\vert+ (1-\alpha) \vert \psi_1 \rangle \langle \psi_1 \vert,
\end{equation}
and
\begin{equation}\label{c1=alpha,c2=0}
\rho(c_1 = \alpha,c_2=0) = \alpha \vert 00 \rangle \langle 00\vert+
(1-\alpha) \vert \psi_1 \rangle \langle \psi_1 \vert,
\end{equation}
where
\begin{equation}\label{psi1}
\vert \psi_1 \rangle  = \frac{1}{\sqrt{2}}( \vert 00 \rangle  +
 \vert 11 \rangle).
\end{equation}
The total correlation $T_2$ is maximal for $c_1 = c_2 =
\frac{\alpha}{2}$.  In this  case, the states of type (\ref{rho}) rewrite as
\begin{equation}\label{c1=c2}
\rho\big(c_1 = \frac{\alpha}{2} , c_2 = \frac{\alpha}{2}\big) = \alpha
\rho_0 + (1-\alpha) \rho_1
\end{equation}
where the states $\rho_1$ and $\rho_0$ are respectively given by
\begin{equation}\label{c1+c2=1}
\rho_1 = \vert \psi_1 \rangle \langle \psi_1 \vert
\end{equation}
where $\vert \psi_1 \rangle $ is given by (\ref{psi1}), and
\begin{equation}\label{c1+c2=0}
\rho_0 = \vert \psi_0\rangle \langle \psi_0 \vert
\end{equation}
where $\vert \psi_0\rangle$ is the  state defined by
$$\vert \psi_0 \rangle = \frac{1}{\sqrt{2}} (\vert 01 \rangle  +  \vert 10 \rangle ).$$
Furthermore, it can be clearly seen (figure 1) that when $ 0 \leq \alpha
\leq 0.5$, the total correlation $T_2$ increases as the parameter
$\alpha$ increases. For instance, for $c_1 = 0.05$ the amount of
classical correlation, present in states (\ref{rho}) with $\alpha =
0.1$, exceeds ones measured by the linear entropy in the states with $\alpha
= 0.2, 0.3, 0.4, 0.5$. The situation is completely different for
$\alpha \geq 0.5$. Indeed, for small values of $c_1$, the total
correlation present in states with $\alpha= 0.6$ is higher than
correlations exhibited by the states with $\alpha = 0.7, 0.8, 0.9$. For
high values of $c_1$ ($c_1 = 0.55 $ for instance), more correlation
is  contained in  the states with $\alpha = 0.9$.

\begin{center}
\includegraphics[width=3in]{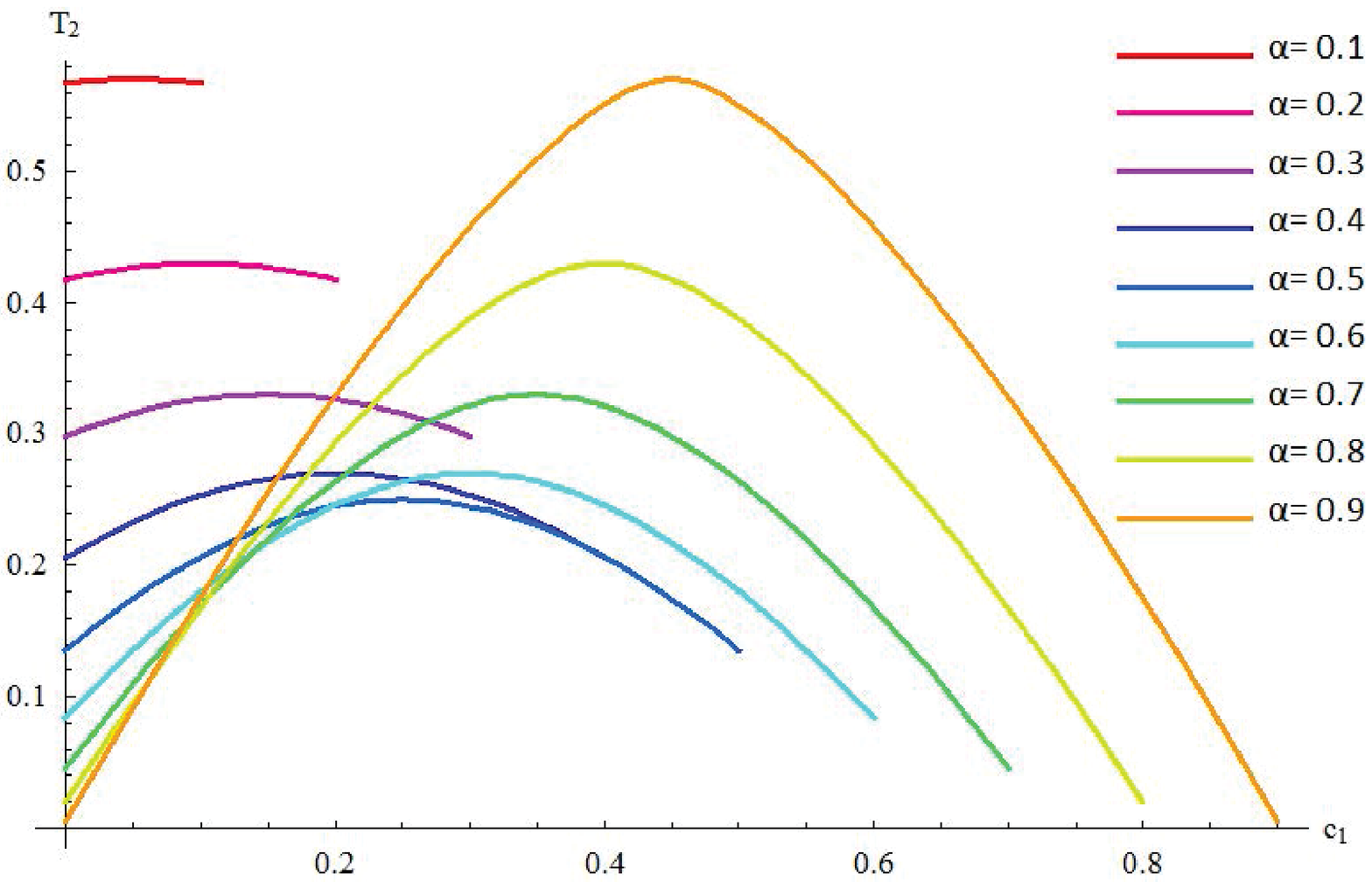}\\
{\bf Figure 1.}  {\sf  Total correlation $T_2$  versus the parameter
$c_1$ for different values of $\alpha = c_1 + c_2$.}
\end{center}

\subsection{Quantum correlation and closest classical state}

\subsubsection{Quantum discord}
The explicit computation of quantum discord, as originally defined in \cite{Vedral-et-al,Ollivier-PRL88-2001}, for an arbitrary
bipartite system is difficult.  Analytical results are known only in a few families of two-qubit states \cite{Luo,Ali,Shi1,Shi2,Adesso,Fanchini1,Fanchini2} (see
also \cite{Daoud1,Daoud2} and references therein). The alternative geometric way,
proposed in \cite{Dakic2010}, quantifies the quantum discord as the minimal Hilbert-Schmidt distance between a given
state $\rho$ and the closest classical states of the form
\begin{equation}\label{cs}
\chi_{\rho}= \sum_{i = 1,2}p_{i}|\psi_{i}\rangle\langle\psi_{i}|\otimes\rho_{i}
\end{equation}
when the measurement is performed on the first subsystem. In
equation (\ref{cs}), $p_i$ is a probability distribution, $\rho_{i}$
is the marginal density matrix of the second subsystem and
$\{|\psi_1\rangle ,|\psi_2\rangle \}$ is an arbitrary orthonormal
vector set. Following the procedure presented in \cite{Dakic2010},  the
explicit expression of the geometric  quantum discord in the state
(\ref{rho}) writes
\begin{equation}
D_\mathrm{g} =\frac{1}{4}\left( R_{11}^2 +  R_{22}^2 +  R_{33}^2 +
R_{03}^2 -\lambda_{\rm {max}}\right) \label{eq:GMQD_original}
\end{equation}
where the correlation elements are given by the expressions ((\ref{R3})-
(\ref{R12})). In (\ref{eq:GMQD_original}), the quantity $\lambda_{\rm{max}}$ is defined by $\lambda_{\rm{max}} = ~\rm{max}~ (\lambda_1 , \lambda_2 , \lambda_3)$
where $\lambda_1$, $\lambda_2$ and $\lambda_3$ denote respectively
the elements of the  diagonal matrix $ K = ~{\rm diag}~ (~ R_{11}^2 , ~ R_{22}^2 , ~R_{33}^2 +  R_{03}^2 ~).$ As we already discussed ,  the
quantum correlation $D_2$ coincides with the geometric measure of
quantum discord. Thus, one gets
\begin{equation}\label{D_2}
D_2 =  D_\mathrm{g} = \frac{1}{4}~ {\rm min}\{ \lambda_1 + \lambda_2
, \lambda_1 + \lambda_3 , \lambda_2 + \lambda_3\}.
\end{equation}
The eigenvalues $\lambda_1$ , $\lambda_2$ and $\lambda_3$, in terms of the parameters $c_1$ and $c_2$,  are
$$ \lambda_1 = \big[ 1 - (\sqrt{c_1} - \sqrt{c_2})^2\big]^2$$
$$ \lambda_2 = \big[ 1 - (\sqrt{c_1} + \sqrt{c_2})^2\big]^2 $$
$$ \lambda_3 = \frac{1}{2}\big[ (3c_1 + c_2 -1)^2 + (c_1 + 3c_2 -1)^2\big].$$
Noticing that  $\lambda_1$ is always greater than $\lambda_2$, we rewrite  the geometric
discord (\ref{D_2}) as
$$D_\mathrm{g} = \frac{1}{4}~ {\rm min}\{ \lambda_1 + \lambda_2 ~,~ \lambda_2 + \lambda_3\} = \frac{1}{4}~ ( {\rm min}(\lambda_1 , \lambda_3) + \lambda_2) .$$
It is simple to verify that the  difference $\lambda_3 - \lambda_1$ is positive when the parameters $c_1$ and $c_2$ satisfy the
 following condition
\begin{equation}\label{condition}
\sqrt{c_1}(2c_1 -1) + \sqrt{c_2}(2c_2 -1) \geq 0.
\end{equation}
Otherwise, we have $\lambda_3
\leq \lambda_1$. Setting
$$ \sqrt{c_1} = e^{-r}\cos \theta,  \quad \sqrt{c_2}= e^{-r}\sin \theta  \quad {\rm with } \quad r \in \mathbb{R} ,~~ 0 \leq \theta \leq \frac{\pi}{2},$$
the condition (\ref{condition}) rewrites
$$e^{-r}(\cos \theta  + \sin \theta)(2e^{-2r}(1- \cos \theta \sin \theta )-1) \geq 0. $$
This inequality  is valid for the variables $r$ and $\theta$ satisfying
$$2e^{-2r}(1- \cos \theta \sin \theta )-1  \geq 0 $$
or equivalently
\begin{equation}\label{condition1}
c_1+c_2 - \sqrt{c_1c_2} \geq \frac{1}{2}
\end{equation}
in terms of the parameters $c_1$ and $c_2$. The set of states of
type (\ref{rho}) can be partitioned  as
$$ \{ \rho \equiv \rho_{c_1,c_2}, ~~ 0 \leq c_1 + c_2 \leq 1 \} = \bigcup_{\alpha \in [0,1] } ~\{ \rho_{\alpha} \equiv \rho_{c_1, \alpha - c_1},  ~~  0 \leq c_1 \leq \alpha \}$$
with $ c_1 + c_2 = \alpha$.
The condition (\ref{condition1}) is satisfied if and only if $\alpha
\geq \frac{1}{2}$. Thus, for a fixed value $\alpha \leq
 \frac{1}{2}$,   the quantity $\lambda_3 - \lambda_1 $ is non
 positive and the geometric measure of quantum discord (\ref{D_2})
 writes
\begin{equation}\label{D_21}
 D_\mathrm{g}=  D_\mathrm{g}^+ = \frac{1}{4}~  (\lambda_2 + \lambda_3).
\end{equation}
For $\alpha \geq
 \frac{1}{2}$, the condition (\ref{condition1}) is satisfied for
$$   0 \leq c_1  \leq \alpha_- \quad  \alpha_+ \leq c_1 \leq  \alpha $$
where
\begin{equation}\label{alpha+-}
\alpha_{\pm} = \frac{\alpha}{2} \pm \frac{1}{2}
\sqrt{(1-\alpha)(3\alpha -1)}.
\end{equation}
In this case, the geometric quantum discord is given by
\begin{equation}\label{D_22}
 D_\mathrm{g} =  D_\mathrm{g}^- = \frac{1}{4}~  (\lambda_1 + \lambda_2).
\end{equation}
Conversely, for $ c_1 \in [ \alpha_- ,  \alpha_+ ] $ the difference
$\lambda_3- \lambda_1$ is negative and the geometric discord reads
\begin{equation}\label{D_22}
 D_\mathrm{g} =  D_\mathrm{g}^+ = \frac{1}{4}~  (\lambda_2 + \lambda_3).
\end{equation}

\subsubsection{Closest classical state}

Now we determine  the explicit form of the closest classical state  to the
state (\ref{rho}). We consider separately the situations
$\lambda_{\rm {max}} = \lambda_1$ and $\lambda_{\rm {max}} =
\lambda_3$. We first treat  the situation where $\lambda_1\leq
\lambda_3$. Following the general procedure developed in
\cite{Dakic2010}, we find that the zero discord (classically
correlated) states  are given by
\begin{equation}\label{classical state first case}
   \chi_{\rho}^{-}=\frac{1}{4}[\sigma_0\otimes\sigma_0+R_{30}\sigma_3\otimes
\sigma_0+
R_{30}\sigma_0\otimes\sigma_{3}+R_{33}\sigma_{3}\otimes\sigma_{3}],
\end{equation}
where the superscript  $^-$ refers to the condition  $\lambda_1-
\lambda_3 \leq 0$. In this case, the pairwise quantum correlation writes
\begin{eqnarray} \label{D-22}
 D^-_2 = S_-(\rho \| \chi_{\rho}^{-})  =   \frac{1}{4}(\lambda_1 + \lambda_2)  =  \frac{1}{4} ( R_{11}^2  + R_{22}^2),
\end{eqnarray}
which can be reexpressed in terms of the parameters $c_1$ and $c_2$ as
$$ D^-_2 \equiv  D^-_\mathrm{g}(\rho) = \frac{1}{4}  \big[1 - (\sqrt{c_1} - \sqrt{c_2})^2\big]^2  + \frac{1}{4}  \big[1 - (\sqrt{c_1} + \sqrt{c_2})^2\big]^2.$$
It is interesting to note that the closest classical state $\chi_{\rho}^{-}$  satisfies
$$ {\rm Tr} \rho \chi_{\rho}^{-}  = {\rm Tr}  \chi_{\rho}^{-^2}$$
traducing that the geometric quantum discord $ D^-_\mathrm{g}$ coincides indeed with the quantum correlation $D^-_2$.
Along the same line of reasoning, one  verifies that for   $\lambda_1> \lambda_3$,  the closest classical state is given by
\begin{eqnarray}\label{classical state second case}
\chi_{\rho}^{+}&=&\frac{1}{4}\Big[
\sigma_0\otimes\sigma_0+R_{03}\sigma_0\otimes\sigma_{3}
   + R_{11}\sigma_1\otimes\sigma_1 \Big],
\end{eqnarray}
where the superscript $^{+}$ refers now to the condition
$\lambda_1- \lambda_3>0$. In this case,  the Hilbert-Schmidt
distance between the density matrix $\rho$ and its closest classical
state $\chi_{\rho}^{+}$  is
\begin{eqnarray}\label{D+22}
  D^+_2 = S_-(\rho \| \chi_{\rho}^{+})  =  \frac{1}{4}(\lambda_2 + \lambda_3)  =  \frac{1}{4} ( R_{22}^2  + R_{03}^2 + R_{33}^2)
\end{eqnarray}
which rewrites also  as
$$  D^+_2  \equiv D^+_\mathrm{g}(\rho) = \frac{1}{4} \bigg[ \big[1 - (\sqrt{c_1} + \sqrt{c_2})^2\big]^2 + \frac{1}{2} \big[(3c_1 + c_2 -1)^2 + (c_1 + 3c_2 -1)^2\big]\bigg].$$
Here again, we notice that the closest classical state  (\ref{classical state second case}) satisfies the following
relation
$$ {\rm Tr} \rho \chi_{\rho}^{+}  = {\rm Tr}  \chi_{\rho}^{+^2}.$$
The behavior of the quantum discord $D_2$, as function of the parameters $c_1$ and $c_2$, is
represented in the figures 2 and 3. Figure 2 gives the amount of
quantum correlations for states with $\alpha \leq \frac{1}{2}$. As it can be inferred
from figure 2, the quantum discord $D_2$ in the states (\ref{rho}) reaches its minimal value
 for $c_1
= \frac{\alpha}{2}$. These "minimally
discordant" states are given by (\ref{c1=c2}). Recall that
the states of type (\ref{c1=c2}) exhibit the maximal amount of total
correlation $T_2$ (see figure 1). In the other hand, the maximal value of quantum
discord  $D_2$ is obtained in the states with $(c_1 = 0, c_2 =\alpha)$ and
$(c_1 = \alpha , c_2 =0)$ which are respectively given  by the
expressions (\ref{c1=0,c2=alpha}) and (\ref{c1=alpha,c2=0}).It is also interesting  to note that in these
"maximally discordant" states the total correlation $T_2$ is minimal (see figure 1). Thus, one concludes that
for the states of type (\ref{rho}) with $c_1+c_2 \leq \frac{1}{2} $, the quantum discord $D_2$ is
maximal (resp. minimal) for states exhibiting a minimal (resp.
maximal) amount of total correlation $T_2$.  The quantum discord $D_2$ and its first derivative
evolves continuously (figure 1). This smooth behavior changes for states with $\alpha \geq \frac{1}{2}$
(figure 3). In fact,  the quantum discord changes suddenly when  $c_1 =
\alpha_-$ and $ c_1 =\alpha_+ $ (  $\alpha_-$ and $ \alpha_+ $ are given by the expressions
(\ref{alpha+-})). This sudden change of quantum discord occurs when
the states $\rho$ (\ref{rho}) have a maximum amount of quantum correlation. The
behavior of quantum discord presents three distinct phases: $0
\leq c_1 \leq \alpha_- $, $\alpha_- \leq c_1 \leq  \alpha_+$ and
$\alpha_- \leq c_1 \leq  \alpha$. The minimal value of quantum
discord $D_2$ is obtained in the intermediate phase ($\alpha_- \leq c_1
\leq  \alpha_+$) for the states given by (\ref{c1=c2}).
\begin{center}
  \includegraphics[width=3in]{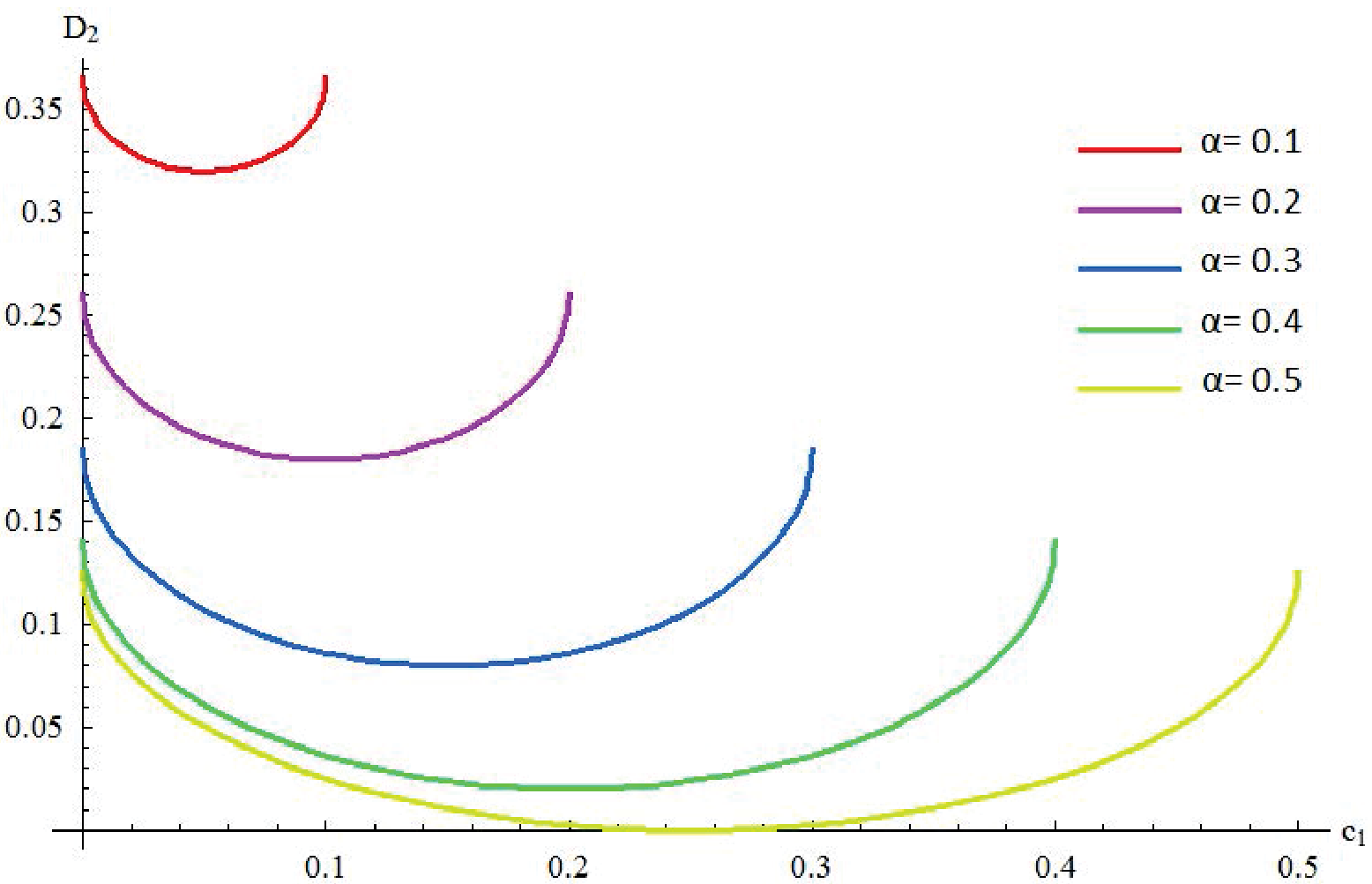}\\
{\bf Figure 2.}  {\sf Quantum discord $D_2\equiv D_{\rm g}$ as
function of the parameter $c_1$ for $\alpha \leq \frac{1}{2}$.}
\end{center}
\begin{center}
  \includegraphics[width=3in]{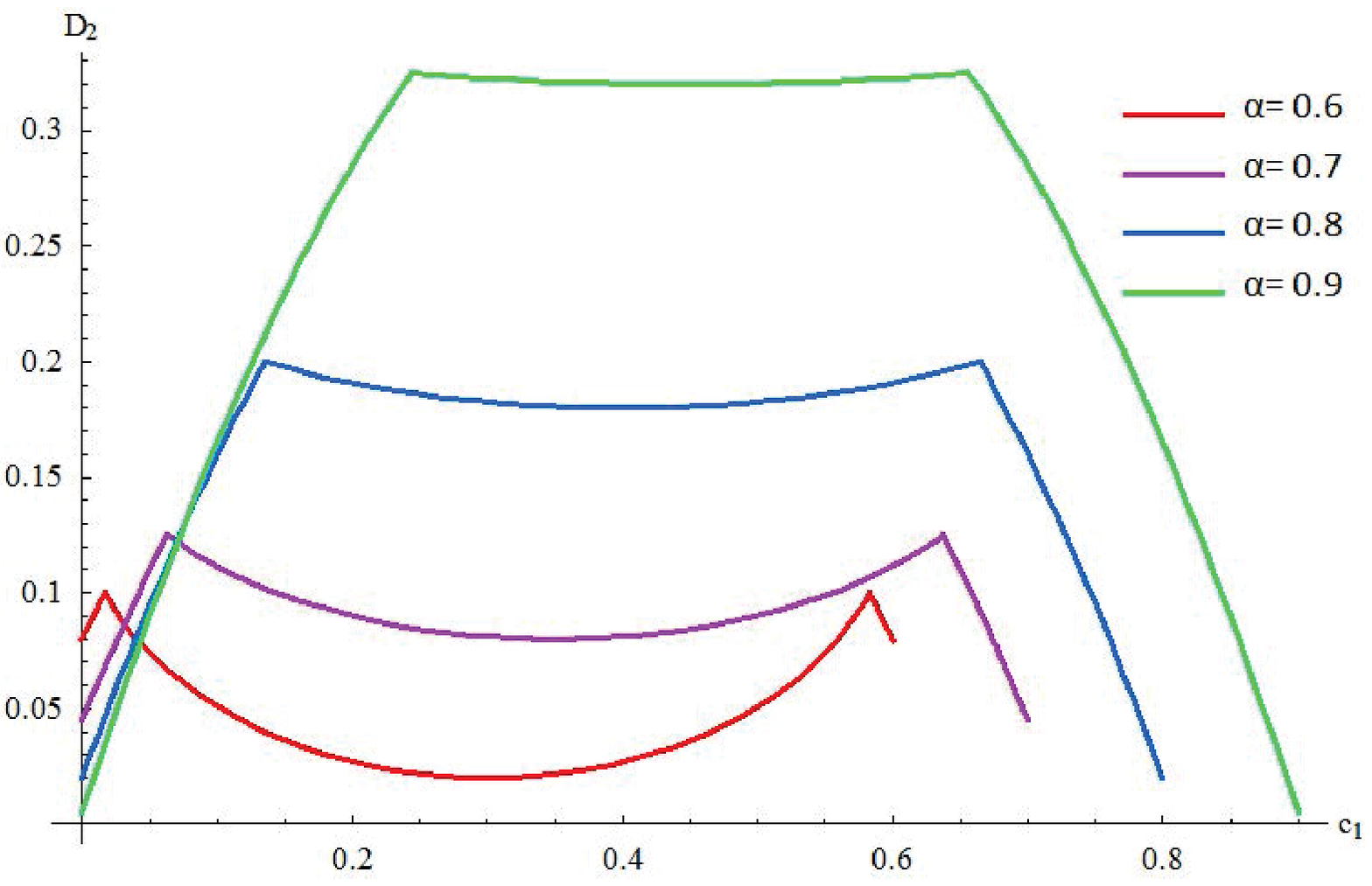}\\
{\bf Figure 3.}  {\sf  Quantum discord $D_2\equiv D_{\rm g}$ as
function of the parameter $c_1$ for $\alpha  \geq \frac{1}{2}$.}
\end{center}
\subsection{ Classical correlations }
The analytical derivation of classical correlations $C_2$
(\ref{T2}) requires the expressions of
 the closest product states to classical states
$\chi_{\rho}^{-}$ (\ref{classical state first case}) and $\chi_{\rho}^{+}$ (\ref{classical state second case}). We discuss first the
situation where the classical state is given by $\chi_{\rho}^{-}$
(\ref{classical state first case}). Remark  that  $\chi_{\rho}^{-}$
possesses parity and exchange symmetries. Thus,  its closest product state can be
obtained using the method leading to  the closest product
state (\ref{pirho}). As result, one gets
\begin{eqnarray}\label{pichi-}
\pi_{{\chi^{-}_{\rho}}} =  \frac{1}{4}\bigg[ \sigma_{0}\otimes \sigma_{0} + a_{3}\sigma_{3}\otimes \sigma_{0}
   +a_{3} \sigma_{0}\otimes\sigma_{3} +  a^2_{3}   \sigma_{3}\otimes\sigma_{3}\bigg].
\end{eqnarray}
which coincides with $\pi_{\rho}$. Subsequently, the classical correlation writes
\begin{eqnarray}\label{c2-}
C^-_2 =\frac{1}{4}\bigg[ 2 ( R^2_{03} - a^2_3 )  + (R^2_{33}-a^4_3)\bigg].
\end{eqnarray}
The determination of the closest classical product to classical
state $\chi_{\rho}^{+}$ (\ref{classical state second case}) is slightly different from the previous
case.  In fact, the state $\chi_{\rho}^{+}$  is invariant under
parity transformation but not invariant  under exchange
symmetry. Accordingly,  the closest classical product states are necessarily of the form
\begin{eqnarray} \label{produitxi}
\pi^+_{\chi_{\rho}} & = & \frac{1}{4}\bigg[ \sigma_{0}\otimes \sigma_{0} +  \alpha_3\sigma_{3}\otimes \sigma_{0}
   +  \beta_3 \sigma_{0}\otimes\sigma_{3} +  \alpha_3 \beta_3 \sigma_{3}\otimes\sigma_{3}\bigg],
\end{eqnarray}
where the variables $\alpha_3$ and $\beta_3$ can be obtained by
minimizing the Hilbert-Schmidt distance between the states
$\chi_{\rho}^{+}$ (\ref{classical state second case}) and
$\pi_{\chi^+_{\rho}}$ (\ref{produitxi}). After some algebra, one shows
$$ \alpha_3 = 0 \qquad \beta_3 = R_{03},$$
and  the closest product state to $\chi_{\rho}^{+}$  writes
\begin{equation}\label{pichi+}
\pi_{\chi^+_{\rho}} =\frac{1}{4}[
\sigma_0\otimes\sigma_0+R_{03}\sigma_0\otimes\sigma_{3}].
\end{equation}
Using the expressions (\ref{classical state second case}),  (\ref{pichi+}) and the definition  (\ref{T2}), the classical correlation reads
\begin{eqnarray}\label{c2+}
C^+_2 =\frac{1}{4} R^2_{11}.
\end{eqnarray}
At this stage, we have the necessary ingredients to calculate explicitly
 the quantity $L_2$ defined by (\ref{T2}).  Indeed, using
the expressions of the closest product $\pi_{\rho}$ (\ref {pirho})
and the closest classical product states $\pi_{{\chi^{-}_{\rho}}}$
(\ref{pichi-}) and $\pi_{{\chi^{+}_{\rho}}}$ (\ref{pichi+}), one has
\begin{eqnarray} \label{L2}
L^-_2 = 0 \qquad           L^+_2 =  \frac{1}{4}\big[ 2 a^2_3   +  a^4_3 - R^2_{03}],
\end{eqnarray}
and one recovers the additivity relation
$$ T_2 + L^{\pm}_2 = D^{\pm}_2 + C^{\pm}_2$$
as excepted.  The figures 4 and 5 give the classical correlations $C_2$  as a function of the parameter $c_1$ for different values
of $\alpha= c_1 + c_2$.  For $\alpha \leq \frac{1}{2}$,
the classical correlation $C_2$ behaves like the total correlation $T_2$ depicted in
figure 1. It is
maximal for states satisfying $c_1 = c_2 = \frac{\alpha}{2}$
(\ref{c1=c2}) and minimal for $(c_1 = 0, c_2 =\alpha)$ (Eq.
(\ref{c1=0,c2=alpha})) and $(c_1 = \alpha , c_2 =0)$ (Eq.
(\ref{c1=alpha,c2=0})). Figure 5 shows a discontinuity of the classical
correlations $C_2$  in the points  $c_1 = \alpha_-$ and $ c_1 =\alpha_+ $
 (Eq.(\ref{alpha+-})). These two particular values are exactly the points  where the quantum discord changes suddenly. This
 discontinuity indicates that the quantity $L_2^+$ is non-vanishing and $C_2 \neq D_2^+ + C_2^+$
 when the parameter $c_1$ ranges from $\alpha_-$ to
 $\alpha_+$. Remark that for
 $c_1=c_2$, the density matrix $\rho$ (\ref{rho}) is a Bell-diagonal state.
 In this case,  we have $R_{03}=0$ and $a_3 = 0$ which implies that the quantity $L_2^+$ is zero. This
shows  that in Bell-diagonal states, the  total correlation $T_2$ is
 exactly the sum of quantum discord $D_2$ and classical
 correlation $C_2$.

\begin{center}
  \includegraphics[width=3in]{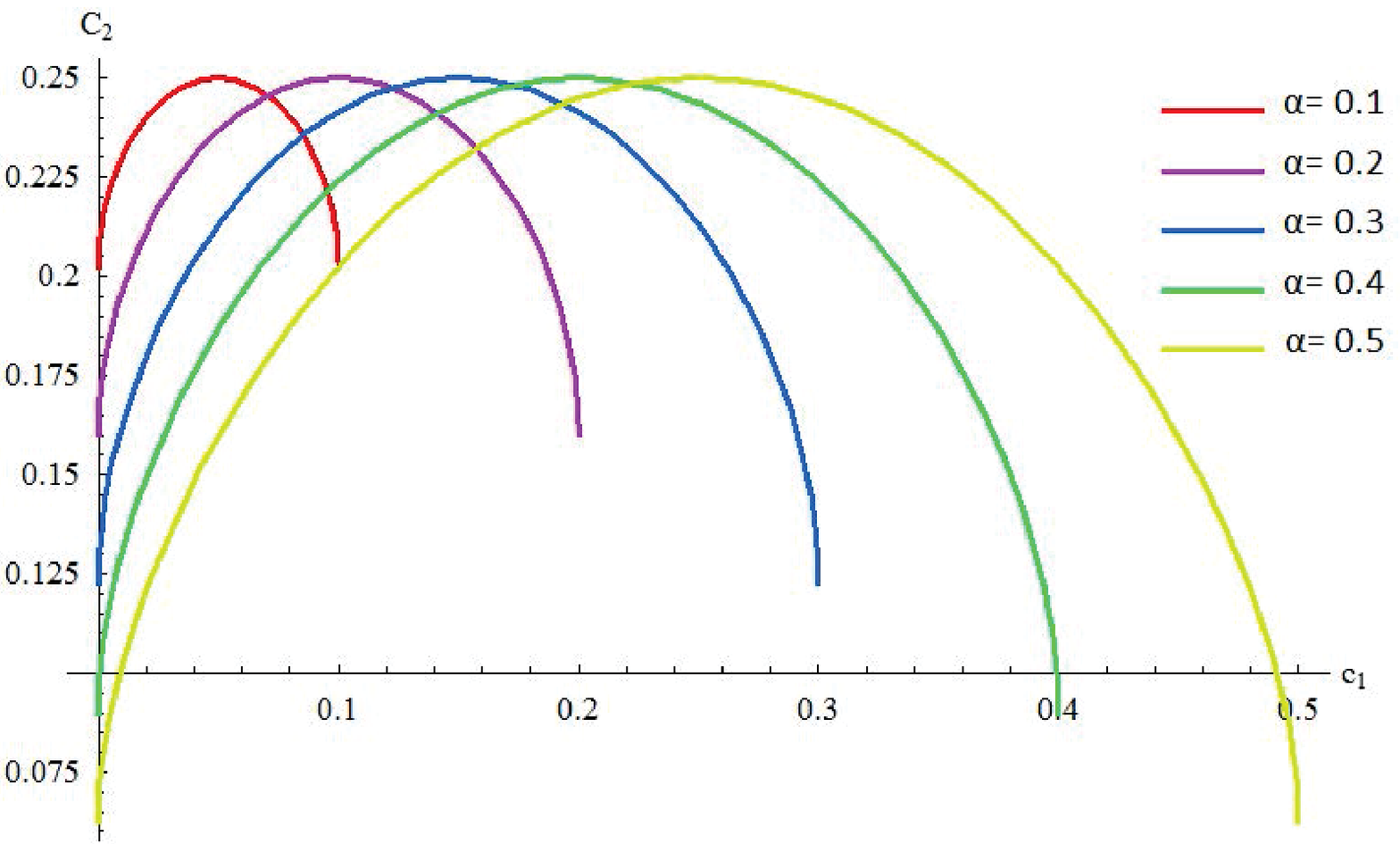}\\
{\bf Figure 4.}  {\sf Classical correlations $C_2$ versus $c_1$ for
$\alpha \leq \frac{1}{2}$.}
\end{center}
\begin{center}
  \includegraphics[width=3in]{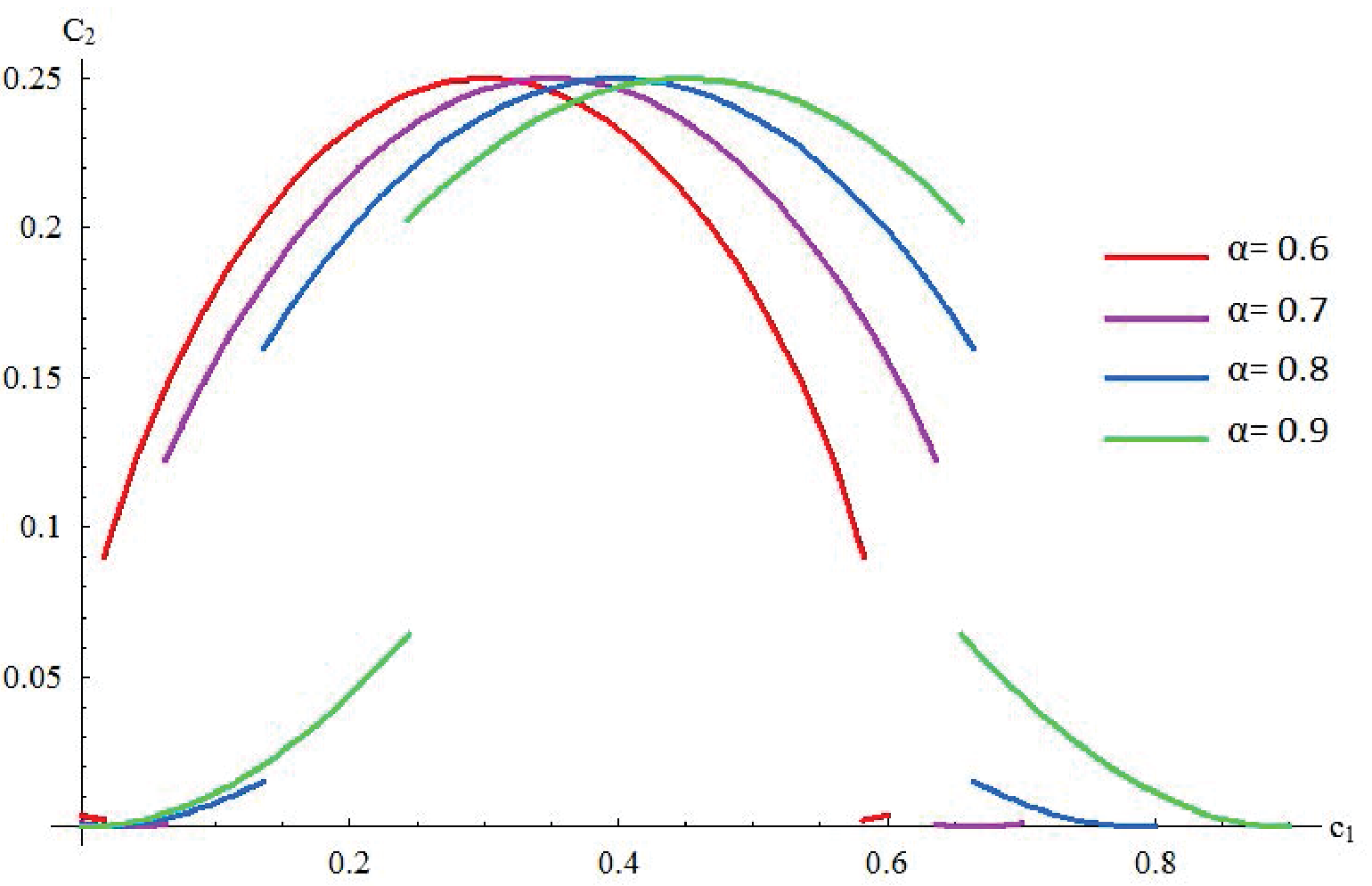}\\
{\bf Figure 5.}  {\sf  Classical correlations $C_2$ versus $c_1$ for
$\alpha \geq \frac{1}{2}$.}
\end{center}

\subsection{Hilbert-Schmidt measures of correlations}

The set of equations (\ref{geometric total-1})  establishes  the relations between the correlation quantifiers
based on relative entropy and geometric correlation quantifiers based on Hilbert-Schmidt distance.
Indeed, using the expressions of closest product
states $\pi_{\rho}$ (\ref{pirho}), one gets
\begin{eqnarray}\label{tr1}
{\rm Tr}\big( \pi_{\rho} (\pi{_\rho} - \rho) \big) = \frac{1}{4}
a^2_3 (R_{33} - a_3^2).
\end{eqnarray}
Similarly, using the closest classical state $\chi_{\rho}^{-}$
({\ref{classical state first case}}) (resp. $\chi_{\rho}^{+}$
(\ref{classical state second case})) and the closest classical
product states $\pi_{{\chi^{-}_{\rho}}}$  (\ref{pichi-}) (resp.
$\pi_{{\chi^{+}_{\rho}}}$ (\ref{pichi+})), one shows
\begin{eqnarray}\label{tr-}
{\rm Tr}\big( \pi_{\chi^-_{\rho}} ( \chi^-_{\rho} - \pi_{\rho}) \big) = \frac{1}{4} a^2_3 (a_3^2 - R_{33})
\end{eqnarray}
and
\begin{eqnarray}\label{tr+}
{\rm Tr}\big( \pi_{\chi^+_{\rho}} ( \chi^+_{\rho} - \pi_{\rho}) \big) =  \frac{1}{4} R_{03} (R_{03} - a_3)
\end{eqnarray}
for $ \lambda_1 \leq \lambda_3$ and $\lambda_1 > \lambda_3$ respectively. Reporting (\ref{tr1}) in (\ref{geometric total-2}) and
using the result (\ref{T2c1c2}), one obtains  the geometric measure of the  total correlation in the state $\rho$
\begin{eqnarray}\label{Tg22}
  T_{g} & = & \frac{1}{4} \bigg[ 2  (R_{30} -  a_3)^2 + R_{11}^2  + R_{22}^2 +  (R_{33} - a_3^2)^2\bigg].
\end{eqnarray}
This result can be also derived from  (\ref{geometric total}) using
the expressions of the closest product state (\ref{pirho}). Analogously, substituting (\ref{tr-}) (resp. (\ref{tr+})) in $C_\mathrm{g}$  (\ref{geometric total-2}) and using
the result (\ref{c2-}) (resp. (\ref{c2+})), one gets
\begin{eqnarray}\label{C-22}
C^-_\mathrm{g}  = \frac{1}{4} \big( 2( R_{30} -  a_3)^2 + (R_{33} -
 a_3^2)^2\big)
\end{eqnarray}
 and
 \begin{eqnarray}\label{C+22}
C^+_\mathrm{g}  = \frac{1}{4} \lambda_1 = \frac{1}{4}  R_{11}^2.
\end{eqnarray}
Inserting the quantities $L_2^{\pm}$  (\ref{L2}) in
 the relations
involving  $L_2^{\pm}$ and $L_\mathrm{g}^{\pm}$ (\ref{geometric total-2}), one has
\begin{eqnarray}\label{L-22}
 L^{-}_\mathrm{g} = 0, \qquad L^{+}_\mathrm{g} =  \frac{a_3^2}{4} \big(1 + a_3^2 + a_3 (a_3^2 - R_{33})^2\big).
\end{eqnarray}
Finally, using the equations (\ref{tr1}), (\ref{tr-}) and
(\ref{tr+}), the quantity $\Delta_g$ (\ref{deltag}) is
given by
 \begin{eqnarray}\label{Delta+22}
 \Delta^-_\mathrm{g} = 0 , \qquad \Delta^+_\mathrm{g} = \frac{1}{2} a^2_3 (R_{33} - a_3^2)
\end{eqnarray}
for $ \lambda_1 \leq \lambda_3$ and $\lambda_1 > \lambda_3$,
respectively. From the results (\ref{D-22}), (\ref{Tg22}),
(\ref{C-22}), (\ref{L-22}) and (\ref{Delta+22}), one verifies that
$$ T^-_\mathrm{g} - D^{-}_\mathrm{g} - C^{-}_\mathrm{g}= 0$$
where $T^-_\mathrm{g}$ denotes the total correlation
$T_\mathrm{g}$ in  the states $\rho$ depending on the parameters  $c_1$ and $c_2$ satisfying the condition  $\lambda_1 \leq
\lambda_3$. In this case, the sum of the quantum correlation  $D^{-}_\mathrm{g}$ and  the classical correlation $C^{-}_\mathrm{g}$
coincides with the total correlation $T^{-}_\mathrm{g}$. This result is no longer valid
in the situation where $\lambda_3 < \lambda_1$. Indeed, from the
equations (\ref{D+22}), (\ref{Tg22}), (\ref{C+22}), (\ref{L-22}) and
(\ref{Delta+22}), we have
$$ T^{+}_\mathrm{g} - D^{+}_\mathrm{g} - C^{+}_\mathrm{g} + L^{+}_\mathrm{g} = \Delta^{+}_\mathrm{g}.$$
Using the equations (\ref{L-22}) and (\ref{Delta+22}), one verifies
\begin{eqnarray}
 \Delta^+_\mathrm{g} -  L^{+}_\mathrm{g} =  - \frac{1}{4} a^2_3 \bigg( a^2_3 +  (a^2_3 + 1 -R_{33})^2\bigg)
\end{eqnarray}
which implies that
$$ T^{+}_\mathrm{g} - D^{+}_\mathrm{g} - C^{+}_\mathrm{g} \leq 0.$$
The  last inequality  becomes an equality for $a_3 = 0$. This solution  is possible when
$R_{30} = 0$ (see equation (\ref{solutiona3})). In this particular case, the state $\rho$ (\ref{rho}) is a
two-qubit state of Bell type. This agrees with the result derived in
\cite{Bellomo2}.



 \section{Concluding remarks}

The concept of relative entropy provides a unified approach to treat
equitably the
different kinds of  correlations existing in multipartite systems \cite{Modi}. In
this approach, total, quantum and classical correlations satisfy a closed additivity
relation. However, the analytical  use of the relative entropy is limited by
the complex optimization procedures  needed in minimizing the distance between a quantum state and its
closest one without the required property.  This constitutes the major drawback  of the
unified view based on the relative entropy. One is therefore forced to consider
other measures. In this sense, Hilbert-Schmidt norm was adopted in \cite{Bellomo1,Bellomo2}
to develop a purely geometric framework  unifying the  geometric variants of total, quantum
and classical correlations. This approach has the advantage of working for a
fairly general class of two qubit systems. But, the geometric
correlations do not satisfy a closed additivity relation
similar to one established in the relative entropy based approach. Motivated
by these reasons, we formulated a scheme reconciliating between the two above mentioned approaches. This
scheme uses a linearized form of relative entropy and provides two major advantages. The first lies in its  analytical simplicity
to determine  the different kinds of correlations. The second advantage concerns
the additivity relation satisfied by the pairwise correlations.
This unified approach has the added merit of being significantly simple in
classifying the different states of bipartite systems according to their
degrees of quantumness.  We also explained how the scheme based on linear
relative entropy can be used systematically to derive the pairwise geometric correlations based on Hilbert-Schmidt
distance.
To exemplify our
analysis, we have considered a special class of two-qubit $X$ states
for which we obtained the analytical expressions of all pairwise
correlations (classical, quantum and total) and the explicit form of
their closest product states, closest classical states and closest
classical product states.

Finally, we mention the general approach, recently proposed in
\cite{Paula2014}, to remove the unexpected ambiguities leading to
multi-valued quantum and classical correlations. Such ambiguities
are essentially due  to degeneracies arising from the optimization
procedures of  distance functions serving as correlations measures.
In this context, we believe that it is worthwhile to examine
 the possible redefinitions of linear
relative entropy quantifiers to avoid such problems. We hope to
treat this issue in a forthcoming work.

\end{document}